\documentclass[preprint2]{aastex}

\slugcomment{}

\shorttitle{Two classes of radio flares in the blazar PKS
0420$-$014}
\shortauthors{Zhou et al.}

\begin{document}


\title{Two classes of radio flares in the blazar PKS 0420$-$014}
\author{J. F. Zhou\altaffilmark{1,2}, X. Y. Hong\altaffilmark{1,2},
 D. R. Jiang\altaffilmark{1,2}, and T. Venturi\altaffilmark{3}}
\altaffiltext{1}{Shanghai Astronomical Observatory, No. 80, Nandan Road,
    Shanghai 200030, China}
\altaffiltext{2}{National Astronomical Observatories, CAS, China}
\altaffiltext{3}{Istituto di Radioastronomia del CNR, Via Gobetti 101,
    I-40129 Bologna, Italy}

\begin{abstract}
The two 5GHz VLBI (Very Long Baseline Interferometry) observations
(1996 June and 1997 November) presented in this paper, combined with
several former
VLBI observations at 8.4GHz and 5GHz, suggest that the radio
flares of the blazar PKS 0420$-$014 can be divided into two classes
according to their geometric origins in 5 or 8.4GHz VLBI maps
and the  properties of light curves. One class of flares, which
we call {\it core flares}, originate from the core. Core flares
have large lags between the light curves at different
frequencies, and will probably lead to the ejection of
new jet components. The other class of flares, which we call {\it jet
flares}, come from jet components. Jet flares vary simultaneously
at different wavelengths, and may due to the Doppler boosting effect
of rotating knots moving along a helical jet. The radio flare
in 1991, accompanied by a simultaneous gamma-ray flare, was
identified as a core flare.
\end{abstract}


\section{INTRODUCTION}
The blazar PKS 0420$-$014 (z=0.915) is active and strongly variable  at
all wavelengths. It is a ROSAT source
\citep{bri94} and
has been identified as an EGRET gamma-ray source \citep{fic94}.

The total flux density of
PKS 0420$-$014 has been monitored extensively at many wavelengths.
The routine monitoring at 4.8, 8.0, and 14.5GHz was carried out by the
26m paraboloid in the University of Michigan \citep{all85}.
The millimeter and submillimeter observations at 22, 37, 90,
150, 230, 270, and 375GHz
 were taken at the 15-m James Clerk Maxwell
Telescope (JCMT), SEST(Swedish-ESO submillimetre Telescope), the
IRAM 30m telescope \citep{ste88, ste92, ste93} and
MRRS(Mets\"{a}hovi Radio Research Station)\citep{ter92}.
 Simultaneous optical and gamma-ray flare was detected in
1992 \citep{wag95}.

The blazar PKS 0420$-$014 has been observed with VLBI
at 43GHz, 8.4GHz, 5GHz and 2.3GHz since 1986.
Five images at 8.4GHz and the results from model fitting are given
in \citet{wag95}
and \citet{fey96}. Images at 5 GHz are presented in
\citet{weh92}, \citet{she97} and \citet{hon99}, who observed the
source in 1986, 1992 and 1995 respectively. These observations show
that the jet components move along a bending trajectory.
Superluminal proper motion was
found in this source \citep{wag95, she97, hon99}.

Some properties of the radio flares were successfully explained by a model
incorporating an adiabatically expanding shocked region \citep{hug89}.
\citet{ste95} investigated the spectral evolution of
the radio flares occurred in 1991 and 1992 in PKS
0420-014.

In this paper we present two 5GHz VLBI observations, carried out
in 1996 June and 1997 November.
An analysis of our data and former VLBI results,
combined with the study of the total flux density variations,
led us to identify the location of the emitting regions of these flares,
which we classify according to the  properties of their light curves.
Throughout this paper, the values $H_0=100 h$
${\rm km}$ ${\rm s^{-1}}$ ${\rm Mpc^{-1}}$ and $q_0=0.5$ will be used.

\section{CLASSIFICATION OF THE RADIO FLARES}
\begin{figure}
\figcaption[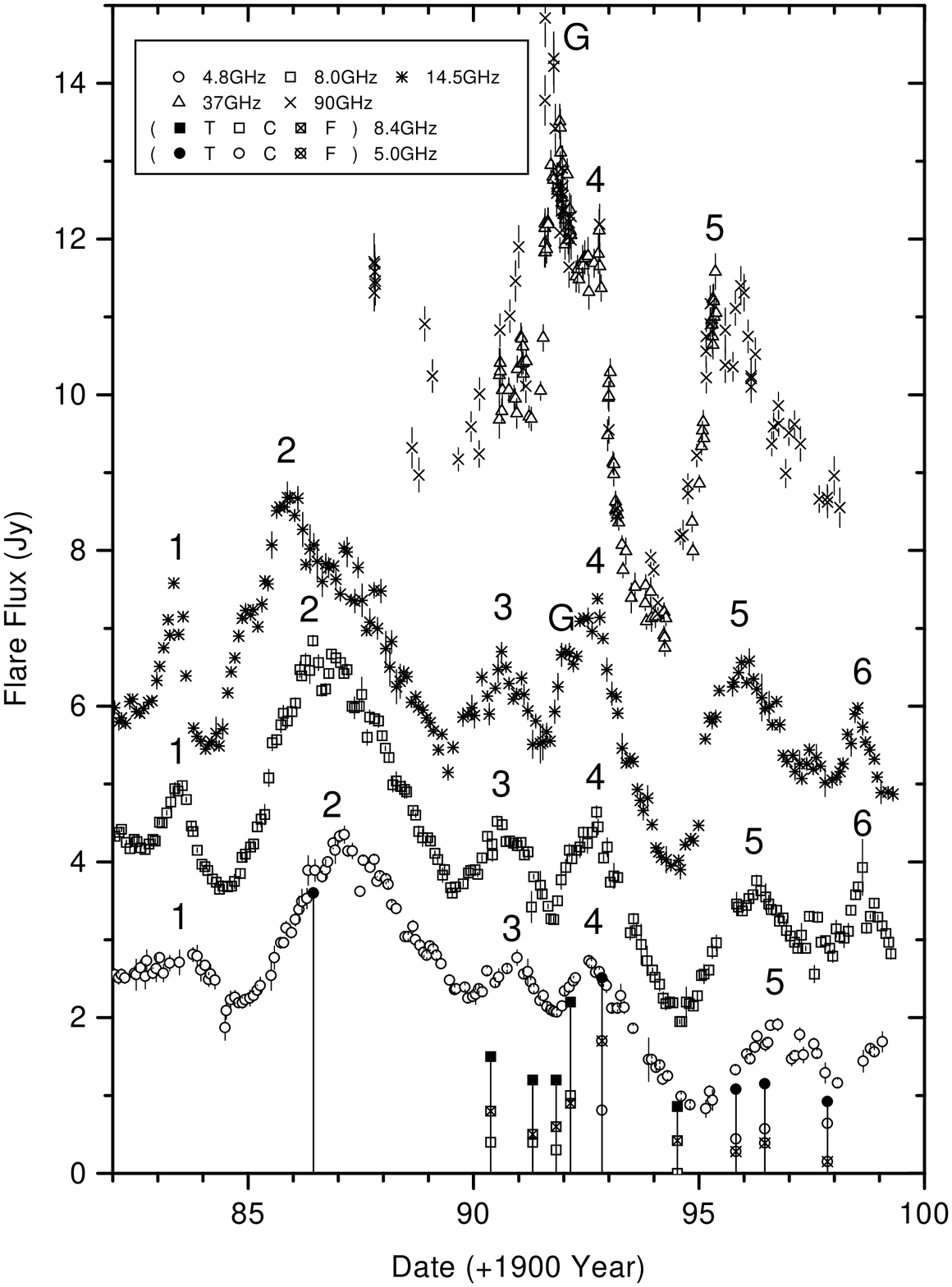]{
Flare flux data of single dish observations and
VLBI observations(T represent the total flare flux
of VLBI structures; C represent the core flare flux; and F represent
the flux of the component A ). Light curves were shifted upwards
0Jy for 4.8, 1Jy for 8.0, 3Jy for 14.5, and 7Jy for 37 and 90GHz
data.
\label{fig1}}
\end{figure}

The long term light curves at 4.8, 8.0, and 14.5GHz are shown in
Figure \ref{fig1}. In order to compare the flare flux density,
we subtracted the quiescent core flux density at each frequency, assuming
a flat quiescent core spectrum.
The quiescent core flux density, about 1.2 Jy, was estimated
from 8.4GHz VLBI observation at epoch 1994.52, when the source was
in its lowest state.
From the data plotted in Figure 1 we can identify
six strong flares(represented by 1, 2, 3, 4, 5, and 6),
which occurred in 1983, 1985, 1990, 1992, 1995,
and 1998. Some minor flares
may also exist, but they are not very significant at these
three frequencies.

We classified the flares according to the time lags of the maximums
at the various frequencies.
Flare 2 and 5 are characterized by large lags between
the peaks of
light curves at different frequencies. We use polynomials to fit
the un-averaged flux data and then determine the peaks of the
light curves(see Table \ref{tbl-0}).
 For  flare 2, the lags $\Delta$t between 14.5 and
8.0 GHz and between 8.0 and 4.8 GHz are
$\Delta$t$^{14.5}_{8.0} = 0.62\pm 0.10$ yr and
$\Delta$t$^{8.0}_{4.8} = 0.52\pm 0.10$  yr respectively.
The corresponding lags for  flare 5 are
$\Delta$t$^{14.5}_{8.0}= 0.36\pm0.10$ yr
and $\Delta$t$^{8.0}_{4.8} = 0.53\pm 0.10$ yr respectively.
We named these as {\it core flares}, or {\it Class I flares}
($\Delta$t$^{14.5}_{8.0}>0.2$ yr).
The remaining strong flares show no or small lags and can
be grouped together. The prototype of this class is  flare 4.
Its light curves reach their peak simultaneously at epoch $1992.6\pm 0.1$.
We named these {\it jet flares} or {\it Class II flares}
($\Delta$t$^{14.5}_{8.0}< 0.2$ yr).

This classification still holds when we add 37 GHz and 90 GHz data
(see Figure \ref{fig1}). Large time lags exist between
the millimeter and centimeter light curves of flare 5,
while flare 4  peaks simultaneously in the whole range of
frequencies from 4.8 GHz to 90 GHz.
A new remarkable feature is a strong millimeter flare
occurred at the end of 1991(represented by G in Figure \ref{fig1}).
The corresponding centimeter flare
was much weaker. Since it shows lags between high and
low frequencies, we classify it as a Class I flare. We will
discuss it in detail in section 4.

\section{VLBI OBSERVATIONS}
Our two 5GHz VLBI observations were carried out
on 1996 June 17, and on 1997 November 07 with EVN(European VLBI Network).
Eight telescopes took part in the 1996 observations, i.e.
Shanghai, Crimea, Noto, Hartebeesthoek(affiliated), Urumqi, Onsala,
WSRT, and Torun.
The nine telescopes participating in the 1997 observations are
Effelsberg, Jodrell Bank, Medicina, Shanghai, Noto,
Hartebeesthoek, Urumqi, WSRT, and Torun.
The recording modes were MkIII mode E (bandwidth 14MHz) and MkIII mode B (bandwidth 28MHz)
for the 1996 and 1997 epochs respectively.
The data were correlated at MPIfR Mk III correlator in Bonn.

\begin{figure}
\figcaption[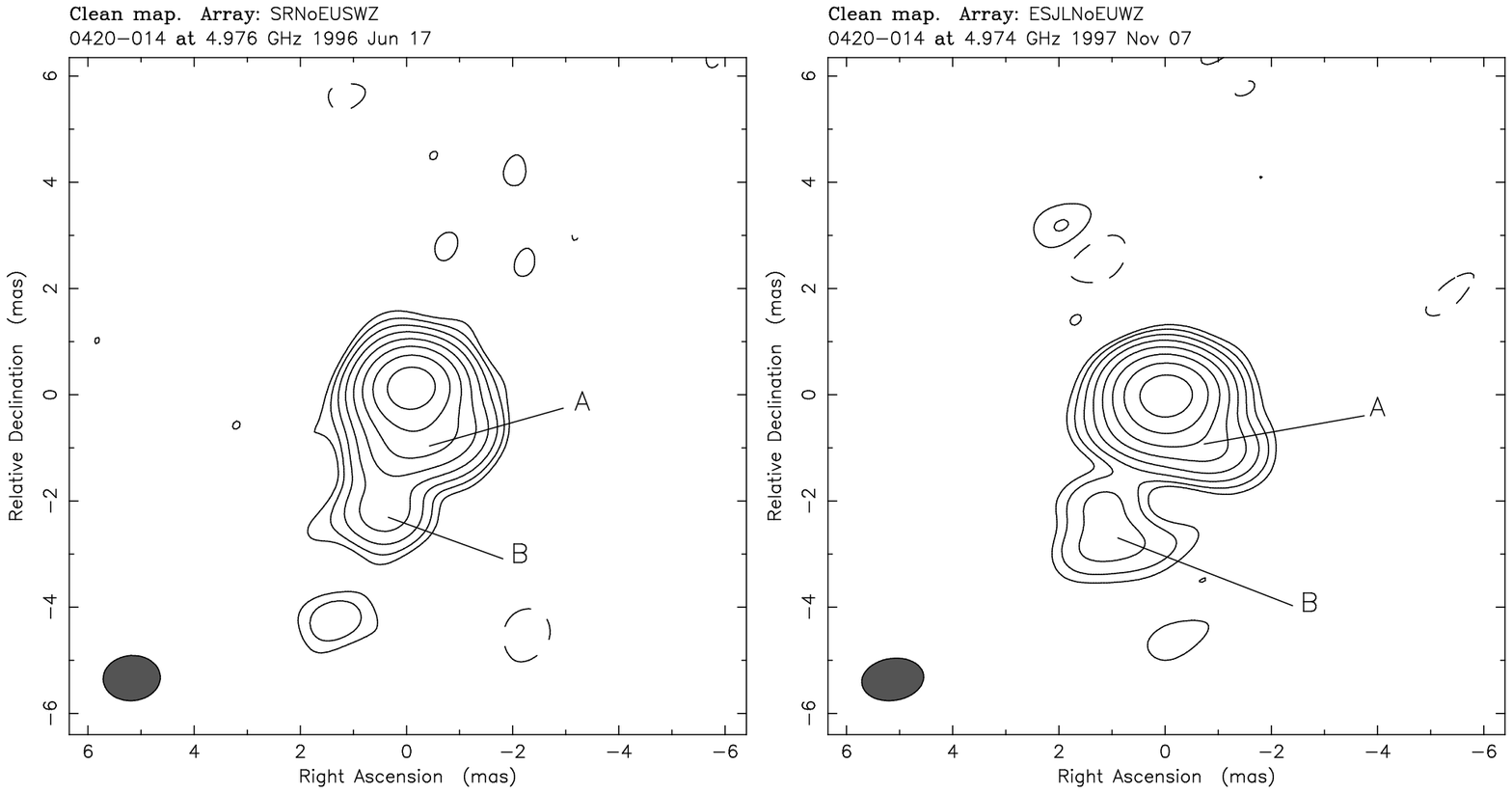]{Two CLEAN maps of 0420-014 in 1996(left) and 1997(right).
A and B indicate
the locations of the first and second jet components.
The contours are -0.5 0.5 1 2 4 8 16 32 and 64 percent of map peaks.\label{fig2}}
\end{figure}

A-priori amplitude calibration and fringe-fitting were carried out
using the standard routines in the AIPS package. Imaging
was done using the DIFMAP difference mapping
software \citep{she94}.
The 1996 and 1997 final images are shown in Figure \ref{fig2}.
For the map of 1996, the noise is 1.68mJy/beam, the map peak is
 1.41Jy/beam, and the FWHM is 1.08x0.85(mas) at $-85.8^\circ$.
For the map of 1997, the noise is 1.85mJy/beam, the map peak is
 1.37Jy/beam, and the  FWHM is 1.17x0.79(mas) at $-81.4^\circ$.
In order to carry out the image analysis,
we used the program `modelfit' in DIFMAP to fit
the final self-calibrated UV data with gaussian components.
The model fitting results are given in Table \ref{tbl-1}.
We estimate that the errors of the parameters are:
$\Delta$Flux = $\pm$ 20$\%$, $\Delta$Radius = $\pm0.1$ mas,
$\Delta$P.A. = $\pm 5^{\circ}$, $\Delta\theta = \pm$ 10$\%$,
$\Delta$Rat. = $\pm$ 15$\%$  and $\Delta$ p.a. = $\pm 20^{\circ}$.

It is clear from Figure \ref{fig2} and Table \ref{tbl-1}
 that the two images
differ considerably.
The size and flux of the core are very similar in the two epochs.
However the position of the first jet component (A) with respect
to the core changed from 1.1 mas to 1.3 mas going from 1996
to 1997. Its position angle also changed, and the flux decreased from
0.39 to 0.15 Jy. The result of our model fitting shows that
component B is stationary.

\begin{figure}
\figcaption[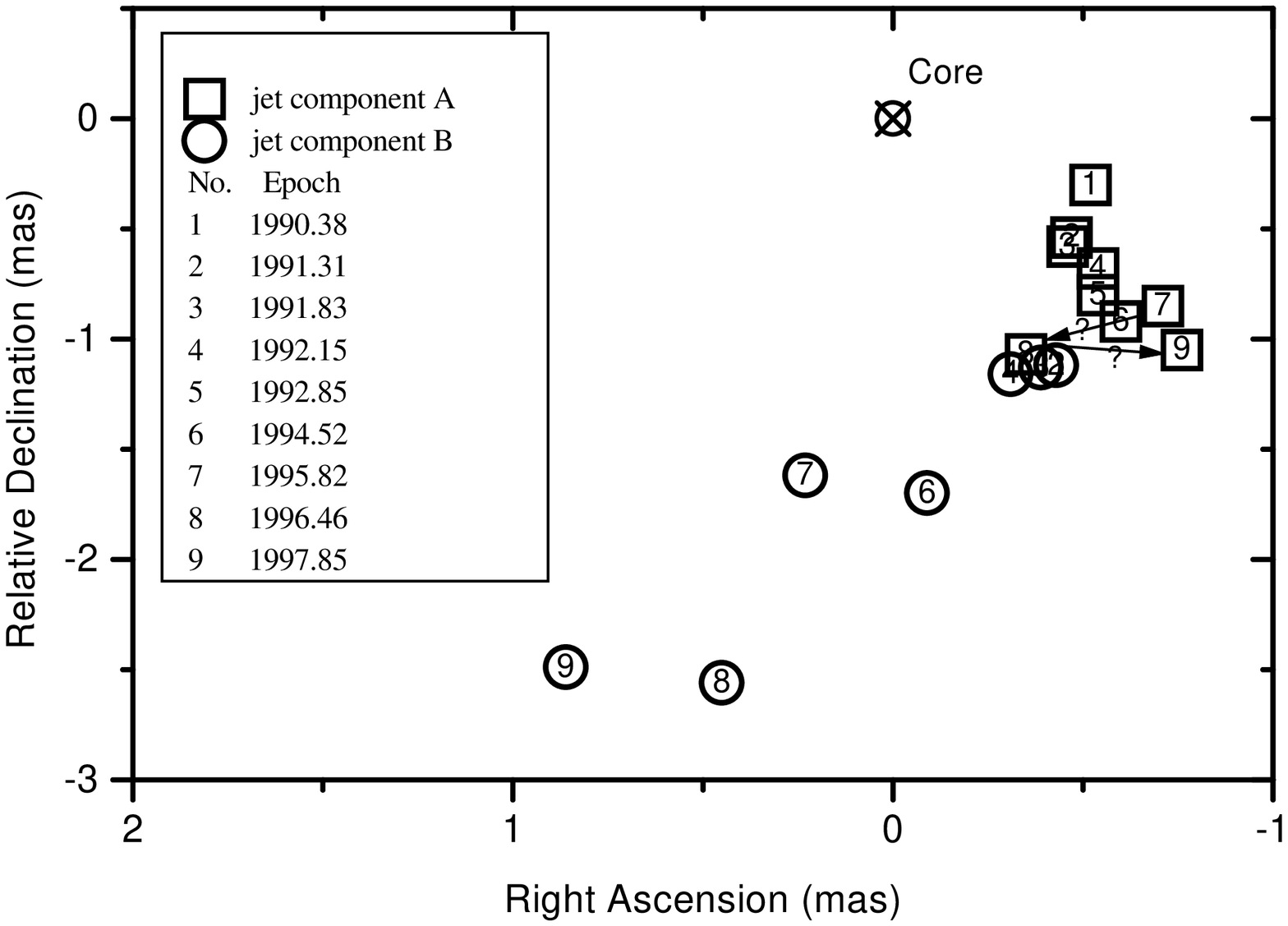]{Relative locations of the jet
components A and B in 0420-014.\label{fig3}}
\end{figure}

In Figure \ref{fig3} we plotted the relative positions of the jet
components with respect to the core from 1990.38 to 1997.85. In
this figure positions 1, 2, 3, 4\citep{wag95}, and 6\citep{fey96}
are from 8.4GHz images; positions 5\citep{she97}, 7\citep{hon99},
8, and 9 are from 5GHz images. The observing epochs are listed in
legend. For each epoch we identified the component closest to the
core with component A in our images, and the more distant
component with B. From Figure 3, it is clear that component A
emerges from the core in the south-west direction. At Radius $\sim
1.3$ mas and P.A. $\approx -145^{\circ}$, the jet bends abruptly
and turns from south-west to south-east, and then slowly curves to
south, which is in agreement with the large scale VLBI map at 18cm
\citep{fey96} and VLA (Very Large Array) map at 20cm
\citep{ant85}.

The position of component A changes smoothly from
1990.38 to 1995.82, suggesting a proper motion
$\mu = 0.12\pm0.02$ mas yr$^{-1}$, which
corresponds to a liner velocity of
$3.1\pm0.5 h^{-1}{\rm c}$. This jet component probably emerged from the
core at $\sim$ 1985.5 when  strong flare 2 visible in Figure
\ref{fig1} began.
While the position of A in 1997.85 is in good agreement with this
proper motion, its location in 1996.46 is difficult to understand,
both because of the large motion from 1995.82 to 1996.46, i.e.
$\mu \sim 0.6$ mas $\rm yr^{-1}$ and because of the abrupt change in
position angle.
It is possible that we mis-identified component A in 1996.46.
We also note that the location of component A in 1996.46 is
in much better agreement with the location of component B at
the epochs labelled 2, 3 and 4.

Another remarkable feature in Figure 3 is
the different kinematical behaviour of component A, within 1.3 mas
from the core, and component B, beyond this distance.
Component B
does not appear continuously along the jet path.
Gaps where the jet is undetected are clearly visible.
This can be easily explained assuming that the jet component (or knot)
moves along a helical jet path and rotates around the jet
axis. Under this assumption the viewing angle of the jet component
changes continuously. When the viewing angle near its minimum value, the
corresponding Doppler factor is large and the component is
visible; when the viewing angle near its maximum value, the
corresponding Doppler factor is small, and the component may
disappear.

In order to locate the origin of the radio flares,  we add
the total VLBI flux density,
the core flux density and the flux
density of component A of ten VLBI observations in
Figure 1. Both the total VLBI flux density and the core flux density
were subtracted by 1.2 Jy of quiescent core flux.
As we can see from Figure 1, the Class 1 flares,
such as the 1985 one, are dominated by a core flux density increase,
while the flux density variations of jet component
are much less and can be neglected.
On the other hand, Class 2 flares are dominated by jet flares. A typical
such flare happened in 1992. Near the peak of this flare, VLBI
observation \citep{she97} showed that the flux of
the component A was 1.7 Jy, much greater than the 0.8 Jy core
flare flux.
The core flare flux density is comparable with the flux
of the component A during the intervals between 1990 and
1992 and between 1995 and 1997.

We conclude that PKS 0420$-$014 is characterized by two different
classes of flares, originating in different source regions:
Class 1 radio flares originate in the core, while Class 2 flares
come from jet components.

\section{DISCUSSION}
Optical depth is essential to
explain the observed phenomena.
In VLBI, 'core' actually refers to the optically thick
inner jet  with flat spectrum, while 'jet' refers to the optically thin
outer jet with steep spectrum. As a new flare component  generates
in the core region,
its emission  will gradually become optically thin from high frequencies
to low frequencies.  As a results, we see the time-lagged
core flares. The jet flares, however, probably happens in the
optically thin jet region.

The classification of radio flares depends on their light
curves as well as VLBI results. However, the core size in
VLBI maps depends on the observing frequency and resolution.
A flare component might be resolved as a jet component in a 43GHz
high resolution VLBI map but still in the core in a
5GHz low resolution VLBI map. So,here, we need to point
out that core flare happens within the core region based on 5GHz or 8.4GHz
VLBI observations, usually with one milliarcsec resolutions.

We propose that jet flares are due to the Doppler boosting effect
of knots moving along a helical jet. All the VLBI observations
available show that the proper motion of the jet component emerged
at 1985.5 is nearly constant. This implies that the variation of
the Lorentz factor and of the viewing angle are small. Small
variation of Lorenz factor would not cause large variation of
flux. However, because the average viewing angle of the jet in PKS
0420$-$014 is small (about $5.5^\circ$ \citet{hon98}), even a
small change in the viewing angle would cause a great change of
Doppler factor as well as the flux of jet component \citep{bla79}.
Furthermore, the Doppler factor is independent of the observing
wavelength, which can naturally explain the simultaneous jet
flares at different wavelengths. The intrinsic variation of the
parameters of jet components  may also cause flux variation, but
cannot fully explain the observed facts.

The flare which took place at the end of 1991 was classified as core flare
on the basis of a 5 months lag between the peak at 90GHz and at
14.5GHz. Furthermore, \citet{kri94} detected
a new jet component in a 43GHz VLBI observation at epoch 1992.40.
This component was about 0.2mas (private communication) away from the
core, and was probably ejected during the late 1991 flare.
However, the core flare in 1991 differs from the other core
flares. It was much stronger at millimeter than at centimeter frequencies;
its flux density increase was very sharply at
the beginning and faded away quickly; last but not least it
was accompanied by a simultaneous gamma-ray flare. Thus,
more work is needed to study the mechanism of this flare.

\acknowledgments
We thank an anonymous referee for his/her helpful comments.
This research has made use of data from the University of Michigan
Radio Astronomy Observatory which is supported by funds from the
University of Michigan. Zhou thanks Merja Tornikoski for generously presenting
the 37GHz and 90GHz SEST observed flux data of PKS 0420-014.
This research has been supported by The Major State Basic Research
Development Program and NNSFC no. 19773019.

\clearpage
\begin{deluxetable}{l|ccccccc}
\tabletypesize{\scriptsize}
\tablecaption{Epochs of the peaks of the lightcurves at different frequencies. \label{tbl-0}}
\tablewidth{0pt}
\tablehead{Freq.&\colhead{1}&\colhead{2}&\colhead{3}&\colhead{4}&\colhead{5}&\colhead{6}&\colhead{G}\\
(GHz) & (yr) & (yr) & (yr) & (yr) & (yr) & (yr) & (yr)}
\startdata
4.8   &  83.54 & 87.16 & 90.89 & 92.54 & 96.71 & x     & x \\
8.0   &  83.40 & 86.64 & 90.67 & 92.60 & 96.24 & 98.59 & x \\
14.5  &  83.35 & 86.02 & 90.61 & 92.60 & 95.88 & 98.52 & 92.04 \\
37    &  x     & x     & x     & 92.63 & x     & x     & 91.76 \\
90    &  x     & x     & x     & x     & 95.62 & x     & 91.64 \\
\enddata
\end{deluxetable}

\clearpage
\begin{deluxetable}{ccrrrrrr}
\tabletypesize{\scriptsize}
\tablecaption{Model fitting results. \label{tbl-1}}
\tablewidth{0pt}
\tablehead{
\colhead{Epoch} & \colhead{Comp.} & \colhead{Flux} &
\colhead{Radius} &
\colhead{P.A.} & \colhead{FWHM} & \colhead{Rat.} &
\colhead{p.a.}\\
 & & (Jy) & (mas) & (deg) & (mas) & & (deg)
}
\startdata
1996.46 & Core &1.77 &0.0 &0.0 &0.62 &0.63 &-20\\
        &  A   &0.39 &1.1 &-162 &1.56 &0.36 &-63 \\
        &  B   &0.18 &2.6 &170 &2.40  &0.01 &84 \\
1997.85 & Core &1.84 &0.0 &0.0 &0.66 &0.69 &28 \\
        &  A   &0.15 &1.3 &-144 & 0.91 &0.01 &81 \\
        &  B   &0.13 &2.6 &161 &2.00 &0.01 &59 \\
\enddata
\tablecomments{Radius and P.A. refer to the relative orientation
w.r.t. the core; FWHM, Rat., and p.a. refer to the average
width, axis ratio and orientation of the individual components.}
\end{deluxetable}

\clearpage

\clearpage

\clearpage

\end{document}